\documentclass[cits]{PoS}

\title{Implications of high-precision spectra of thermonuclear X-ray bursts for
determining neutron star masses and radii}

\ShortTitle{Neutron star masses and radii from X-ray bursts}

\author{\speaker{M. Coleman Miller}\\
        University of Maryland, Department of Astronomy and Joint
        Space-Science Institute, College Park, MD  20742-2421, USA\\
        E-mail: \email{miller@astro.umd.edu}}

\author{Stratos Boutloukos\\
        Theoretical Astrophysics, University of T\"ubingen, Auf der
        Morgenstelle 10, 72076, Germany\\
        E-mail: \email{stratos@tat.physik.uni-tuebingen.de}}

\author{Ka Ho Lo\\
        Center for Theoretical Astrophysics and Department of Physics,
        University of Illinois at Urbana-Champaign, 1110 West Green Street,
        Urbana, IL 61801-3080, USA\\
        E-mail: \email{kaholo2@illinois.edu}}

\author{Frederick K. Lamb\\
        Center for Theoretical Astrophysics and Department of Physics,
        University of Illinois at Urbana-Champaign, 1110 West Green
        Street, Urbana, IL 61801-3080, USA and
        Department of Astronomy, University of Illinois at Urbana-Champaign, 1002
        West Green Street, Urbana, IL 61801-3074, USA\\
        E-mail: \email{fkl@illinois.edu}}

\abstract{X-ray burst spectra have long been used to estimate neutron star
masses and radii. These estimates assumed that burst spectra are
accurately described by the model atmosphere spectra developed
over the last three decades. We compared RXTE data from a superburst
with these spectra and found that the spectra predicted by previously
published model atmospheres are strongly inconsistent with these 
high-precision measurements.  In contrast, a simple Bose-Einstein
spectrum is fully consistent with the data, as are recently
published model atmosphere spectra. We discuss the implications of our
results for determinations of neutron star masses and radii via
constraints on their surface gravity and redshift, as originally
suggested by Majczyna and Madej.}

\FullConference{Fast X-ray timing and spectroscopy at extreme count rates: 
Science with the HTRS on the International X-ray Observatory - HTRS2011,\\
		February 7-11, 2011\\
		Champ\'ery, Switzerland}

\begin{document}

\section{Introduction}

A few years after the discovery of thermonuclear X-ray bursts from
accreting neutron stars, Jan van Paradijs proposed a method for using 
observations of thermonuclear X-ray bursts to constrain both the masses 
and radii of the stars and hence to provide key information
on the properties of cold high-density matter \cite{vP79}.
In brief, the argument was that (1)~if the luminosity of a source
during the so-called touchdown phase of photospheric radius expansion
bursts was the Eddington luminosity of the neutron star, and (2)~if
during the cooling phase of the bursts the entire surface of the star
emits uniformly, then a combination of the observed touchdown flux
and area normalization plus knowledge of the distance to the source
and the composition of its atmosphere suffices to determine the star's
mass and radius.

The first applications of this method yielded puzzling results.
Burst spectra are very close to Planck spectra, but the fitted
Planck temperatures are commonly $kT_{\rm fit}\sim 3$~keV at the
peaks of bursts, which is higher than is possible if the atmosphere
is purely gravitationally confined \cite{mars82}.  
Also, in many cases application of this method leads to estimates of the
stellar radius that are implausibly small ($<5$~km).  It was then pointed
out that although the {\it shape} of the spectrum may be qualitatively
similar to a Planck spectrum,
atmospheric opacity effects can shift the peak of the spectrum
so that Planck fits of X-ray data yield a fitted temperature that can
be up to $\sim 2$ times the surface effective temperature.  It has
been largely accepted that such models describe the spectra correctly,
but prior to our work no comparison had been made with data that
are capable of distinguishing between simple Planck or Bose-Einstein
spectra and model atmosphere spectra; the differences are subtle,
and require data taken with the best available instrument (the
{\it Rossi} X-ray Timing Explorer Proportional Counter Array [RXTE PCA])
from long bursts that maintain steady spectra for tens of seconds
as opposed to the tenths of a second that are usual for typical bursts.

Here we describe and elaborate on the comparisons we first reported
in \cite{bout10}.  We find, surprisingly, that although  a simple
Bose-Einstein function fits the highest-precision single spectra 
available in the RXTE archive, the most commonly used atmospheric
spectral models are inconsistent with such spectra.  This calls
into question inferences made using these models.  In Section~2 we
give an overview of the principles behind atmospheric spectral models and
why they shift the spectral peak.  In Section~3 we discuss our 
comparisons with RXTE data and what they imply.  Finally, in Section~4 we
discuss ongoing work in which we compare new atmospheric
spectral models with the data, and in particular the indications that 
these models may fit long data sets better than
Bose-Einstein models.  We discuss the implications
of such fits, particularly that they may yield 
constraints on the mass and radius via joint constraints on the
surface gravity and redshift, but caution that the
approximations in current models do not yet allow us to draw robust
conclusions.

\section{The Principles of Burst Model Atmosphere Spectra}

In the past three decades, many groups have calculated model atmosphere
spectra relevant for bursts (e.g., \cite{lond84,made04,majc05}).  
The high temperatures of the bursts
mean that unless the atmosphere has unexpectedly large metallicity, 
atoms will be fully ionized.  The only opacity sources are then
free-free absorption (important at sufficiently low photon energies)
and Compton scattering (expected to dominate over most or all of the
observed PCA energy range).  

\begin{figure}
\begin{center}
\includegraphics[width=0.6\textwidth]{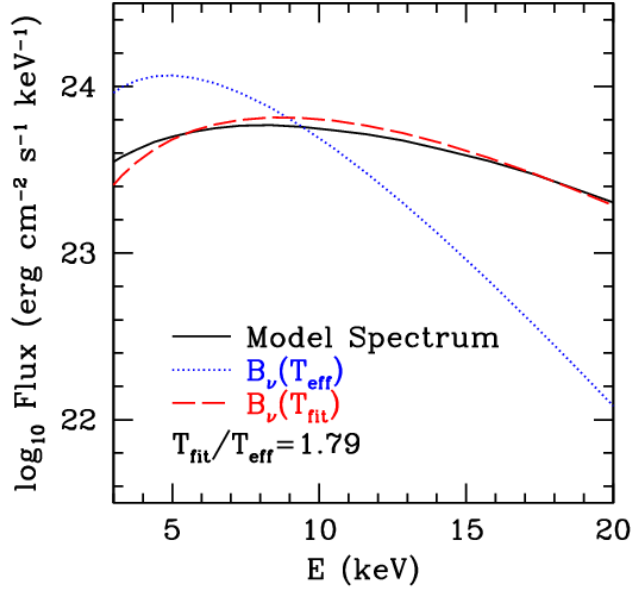}
\end{center}
\vspace{-0.8cm}
\caption{Illustrative figure showing the upward shift in the peak of X-ray
burst spectra produced by model atmospheres. The solid line shows
a model spectrum from \cite{majc05}, the dashed line is the best
fit of the Planck function to the spectrum, with an adjustable normalization to
describe the reduction of the emergent flux caused by scattering, and
the dotted line is the Planck spectrum at the effective temperature.
For this case, where the flux is $\sim 80$\% of 
Eddington, the best-fit Planck temperature has a temperature 
$\sim 1.8$ times the effective temperature.  This is the main reason the 
dotted Planck spectrum differs from the best-fit Planck spectrum.  
Although the shape of the
model atmosphere spectrum is close to the shape of a Planck spectrum,
deviations are evident at low energies and at high energies.}
\label{tcorr}
\end{figure}

In an idealized situation where the only opacity is
energy-independent scattering, we can understand the shift in
the peak of the spectrum caused by the scattering using a simple thought
experiment.  Suppose that the atmosphere has a net surface radiative
flux $F=\sigma T_{\rm eff}^4$, where $\sigma$ is the Stefan-Boltzmann
constant and $T_{\rm eff}$ is the effective temperature.  Suppose also
that it is in complete thermal balance and hence emits a Planck
spectrum, but that on top of the atmosphere is a scattering layer that
lets an energy-independent fraction $0<f<1$ of the photons through,
the rest being reflected and rethermalized in the atmosphere.  
Because the flux $F$ must
emerge, the atmosphere heats up, still in thermal equilibrium, to a
temperature $T_{\rm fit}=f^{-1/4}T_{\rm eff}$.   The net flux is still
$F=f\sigma(f^{-1/4}T_{\rm eff})^4=\sigma T_{\rm eff}^4$, so the
effective temperature ({\it defined} as $T_{\rm
eff}=(F/\sigma)^{1/4}$) is unchanged and the emergent spectrum is still
a perfect Planck spectrum, but its temperature is $f^{-1/4}$
times the effective temperature.  Figure~\ref{tcorr} shows a typical example
\cite{majc05} of how model atmospheres shift the
peak of the spectrum upward.

Figure~\ref{tcorr} also shows how model atmosphere spectra typically 
deviate from the Planck spectrum that best fits them.
The deviation at low energies is caused
primarily by the energy-dependence of the free-free opacity whereas
the deviation at high energies is due primarily to the
energy-dependent Klein-Nishina correction to the Thomson scattering
cross section. Thus although the model atmosphere spectra have
shapes that are close to the shape of a Planck spectrum, there are
deviations that can in principle be observed. 

\section{Comparison of Models with Data}

Prior to our work in \cite{bout10}, very few comparisons had been
made of model spectra with burst data, and none used data with enough
counts to distinguish between qualitatively different models (e.g.,
Planck spectra fitted at least as well as model atmosphere spectra in
the work of \cite{fost86}).  It is therefore critical to
use long stretches of data taken with the
RXTE PCA during intervals when the temperature is nearly constant. 

Most thermonuclear X-ray bursts last only a few seconds, during which
time the temperature changes rapidly enough that a single-temperature
fit is only appropriate for data segments shorter than a few tenths of
a second.  However, we found that around the peak of the superburst
from 4U~1820--30 (see \cite{stro02}), there was a 64-second segment
with $\sim\,800,000$ counts that had a nearly constant temperature. 
This is the most precise available data set.  We note that although
the nuclear processes in superbursts and canonical bursts are
different, their atmospheric processes are the same and hence for the
purpose of spectral fitting this is a representative data set.  We
also note that in the later portions of this burst, high time
resolution data show no evidence that the spectrum changes on time
scales $<10$~s, supporting our expectation that the time scale of
variability is much longer in superbursts than in canonical bursts.

Our first comparison was with a Bose-Einstein spectral model, in which
the continuum is
\begin{equation}
F(E,T)\propto E^3/\left[\exp((E-\mu)/kT)-1\right]\; .
\end{equation}
Here $E$ is the photon energy, $T$ is the temperature, and $\mu<0$ is
the chemical potential.  This spectrum, which generalizes and is more
physically realizable than a Planck spectrum, is the equilibrium spectrum
for fully saturated Comptonization; it could thus be a reasonable
approximation to the spectrum produced in a scattering-dominated atmosphere
\cite{bout10}.  In addition to the 
continuum component, we follow \cite{stro02} in adding as
additional components that originate far from the star a zero-redshift
iron emission line, an edge, and photoelectric absorption.

\begin{figure}
\begin{center}
\includegraphics[width=0.6\textwidth]{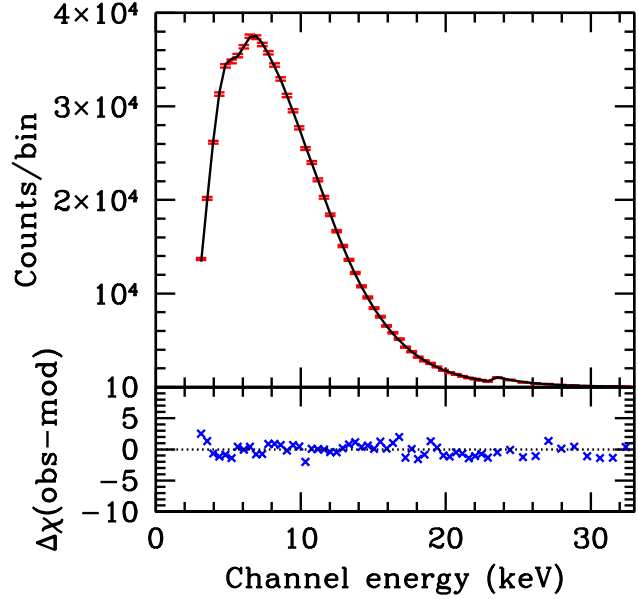}
\end{center}
\vspace{-0.8cm}
\caption{Fit of a model with a Bose-Einstein continuum plus a zero-redshift
iron line and edge and photoelectric absorption to $\sim$800,000 counts
of data near the peak of a superburst from 4U~1820--30.  The top panel
shows the count data (shown with error bars representing the statistical
uncertainties in the data) and the fit, shown by a solid line.  The
bottom panel shows the residuals.  Contrary to our initial expectations,
the fit is superb.  Figure adapted from \cite{bout10}.}
\label{befit}
\end{figure}

We show the result in Figure~\ref{befit}.  Remarkably, the simple
Bose-Einstein form fits this $\sim$800,000 count spectrum extremely
well, with $\chi^2$/dof=55.8/50 over the 3--32~keV range of our fit.
The best-fit temperature and chemical potential are $kT=2.85$~keV and 
$\mu=-0.76$~keV.  The data here are from where the flux measured with
RXTE is $\sim 90$\% of the peak flux of the burst, but we also find
good fits to data at 100\%, 80\%, and 25\% of the peak measured flux.
The excellent fit of the Bose-Einstein shape is therefore not confined
to the peak.

The high quality of this fit suggests challenges for spectral modelers.
In particular, two questions emerge: why are the spectra so close to
Bose-Einstein, and why is the magnitude of the chemical potential much
less than $kT$?  To elaborate on the latter point: if there were a
significant deficit of photons compared to what would be expected for
a Planck spectrum at $kT$, then $\mu<-kT$, so $|\mu|\ll kT$ implies that the
supply of photons is close to what is needed to fill a Planck spectrum.
Ongoing work by Fred Lamb and Ka Ho Lo suggests that these requirements
can be met in extended atmospheres with appropriate densities (low enough
that scattering dominates, but high enough that photons can be supplied
at the required rate).  It is an open question whether these requirements
are met in realistic models.

Although a Bose-Einstein model fits the highest-precision PCA data
well, the implications are difficult to establish with certainty. This
is because, as we indicated earlier, Thomson scattering in the outer
atmosphere can in principle impose a large dilution factor without
causing any deviation from a nearly-perfect Planck or Bose-Einstein
spectrum established at larger optical depths. In this case the
efficiency $f$ of the emission can be less than unity by a 
significant factor. If the emission efficiency is high, the spectrum
we have measured implies that the surface radiative flux is
significantly super-Eddington and extra confinement is required (e.g.,
\cite{bout10} explored confinement by a tangled magnetic field
generated during bursts).  If instead the efficiency is low, the
surface radiative flux could be sub-Eddington.  For conventional,
gravitationally-confined atmospheres to be favored would require the
spectra they predict to fit much better than a Bose-Einstein spectrum,
but this is not possible for single data segments because our fits of
Bose-Einstein models to such segments yield  $\chi^2/{\rm dof}\sim
1$.  It is, nonetheless, important to determine whether published model
atmosphere spectra also yield $\chi^2/{\rm dof}\sim 1$, 
because such models are only viable if this is the case.

\begin{figure}
\begin{center}
\includegraphics[width=0.6\textwidth]{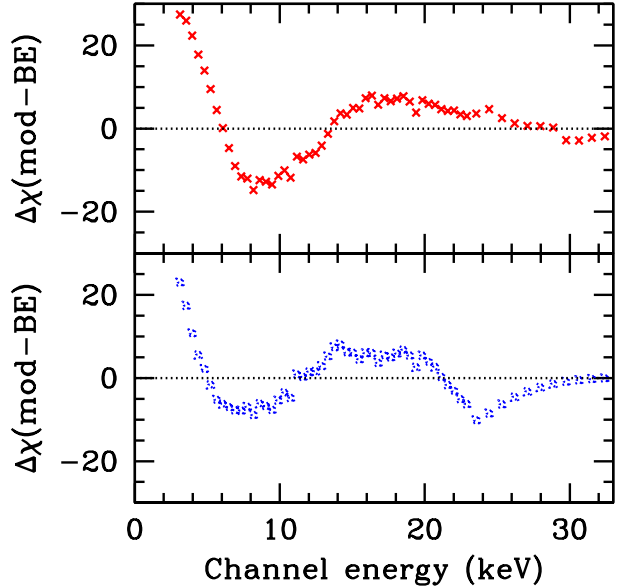}
\end{center}
\vspace{-0.8cm}
\caption{Fit of a Bose-Einstein continuum spectrum to continuum data
with $\sim\,800,000$ counts synthesized using (top panel) a H/He composition,
$\log_{10} (g/{\rm cm~s}^{-2})=14.8$, $T_{\rm eff}=3\times 10^7$~K
model spectrum from \cite{made04} ($F=0.8~F_{\rm Edd}$ for this
spectrum) and (bottom panel) a solar composition,
$\log_{10} (g/{\rm cm~s}^{-2})=14.3$,  $T_{\rm eff}=2\times 10^7$~K
model spectrum from \cite{majc05} ($F=0.5~F_{\rm Edd}$).
Clearly, the predicted model spectra are very different
in form from the Bose-Einstein shape, and hence from observed spectra.
Caution is therefore appropriate in drawing inferences about stellar
masses and radii using these models. Figure adapted from \cite{bout10}.}
\label{model}
\end{figure}

We show such a comparison in Figure~\ref{model}, where we compare
representative models from \cite{made04,majc05} with the same 64
seconds of data from the 4U~1820--30 superburst that we used previously.
A direct fit of the data is not possible, because the available grids of 
these models are not fine enough and the relevant composition
(pure helium) is not computed.  Therefore, as we did in \cite{bout10},
we compare the {\it shape} of the model spectra with the {\it shape}
of Bose-Einstein spectra.  That is, starting from our observation that
the observed spectra are very close to Bose-Einstein in form, we
produce synthetic RXTE data using the model spectra and fit those data with
a Bose-Einstein model.  As can be seen from Figure~\ref{model}, there are
strong and systematic deviations between these shapes. These
deviations are similar for different compositions (H/He with no
metals versus a solar composition), surface gravities  ($\log_{10}
(g/{\rm cm~s}^{-2})=14.8$ versus 14.3), effective  temperatures
($T_{\rm eff}=3\times 10^7$~K versus $2\times 10^7$~K), and surface
radiative fluxes relative to the Eddington flux ($F=0.8~F_{\rm Edd}$
versus $0.5~F_{\rm Edd}$).  We conclude that the spectral shape
predicted by these models is significantly different from what is
observed. We also found this to be true in a later segment of data
where the observed flux was $\sim 50$\% of the maximum, versus $\sim
90$\% in our primary data set.

Given that use of spectral models that are inconsistent with the best data
may introduce systematic errors in estimates of neutron star masses and
radii, caution seems warranted.  An additional indicator of possible
biases in such estimates was mentioned briefly by \cite{guve10},
and in more detail by \cite{stei10}.  
When the standard assumptions of Eddington luminosity at touchdown
and full-surface uniform emission in the burst tail are employed along
with the best measurements of quantities such as the distance, touchdown
flux, and area normalization, the derived mass and radius are not real
but are instead complex quantities, an obvious impossibility.  Indeed,
\cite{stei10} find that only a fraction $1.5\times 10^{-8}$
of the prior probability distribution of these quantities employed by
\cite{guve10} allow a solution for 4U~1820--30.  
Such a small allowed region in parameter space produces small error bars
on the mass and radius, but may indicate that the assumptions on which the
analysis is based are incorrect.

Although previously published models differ strongly from the most precise data,
more recent models show promise of much better fits.  We discuss these
in the next section.

\section{More Recent Models and Future Directions}

Recently, new burst model atmosphere spectra have been calculated 
\cite{sule10}.  These spectra were computed using an
approximate scattering integral (e.g., the Fokker-Planck
approximation was made), but a fine enough grid was constructed
with enough different compositions (including pure helium) that
they could be fit directly to PCA data.  Our preliminary results using
these spectra are very encouraging; for example, a pure helium
atmosphere with $F=0.95~F_{\rm Edd}$ fits our 64-second segment of data
with $\chi^2/{\rm dof}=42.3/48$.  This is better, but not significantly, than
the best fit of Bose-Einstein spectra to the same data.  These new models
provide comparably
good fits to data later in the burst, when the observed flux is half the
maximum and previously published models still have shapes strongly
discrepant with what is observed. 

This is encouraging, and one might at first imagine that this would
allow us to apply the van Paradijs \cite{vP79} method using the models 
reported in \cite{sule10} or new ones computed without some of the current 
approximations.  Unfortunately, this appears not
to be the case.  We fit 102 consecutive 16-second data segments near the
beginning of the 4U~1820--30 superburst (but after apparent touchdown)
using the models from \cite{sule10}, and found that even when we fixed
the surface gravity and surface redshift (hence fixing the radius of the 
emitting surface) the inferred size of the emitting area changes 
systematically by $\sim 20$\% over the data segments.
One might wonder whether the whole star is, in fact, emitting but the
photospheric or thermalization radius is changing. But a change of the
amount observed would require a surface
radiative flux very close to Eddington to  achieve the necessary large
scale height, and such fluxes are highly inconsistent with the
observed spectra.  Instead, it appears that the fraction of the
surface that emits changes systematically, in conflict with the
standard simplifying assumption of the van Paradijs method.

The encouragingly good fits using the models from \cite{sule10}
do suggest an alternative method for determining the mass and radius, 
originally suggested in
\cite{majc05b}.  In addition to composition and surface radiative flux,
the surface gravity is a parameter in the models and to relate the
surface spectrum to what we see at infinity we must also include the
surface redshift in the fit.  The surface gravity $g$ and surface redshift 
$z$ depend
differently on the gravitational mass $M$ and circumferential radius $R$;
for example, for a nonrotating star whose exterior spacetime is therefore
Schwarzschild, $1+z=(1-2GM/Rc^2)^{-1/2}$ and $g=(GM/R^2)(1+z)$.  
Inverting then gives us
\begin{equation}
R=(c^2/2g)(1-1/(1+z)^2)(1+z)\quad{\rm and}\  M=(Rc^2/2G)(1-1/(1+z)^2)\;,
\end{equation}
where $c$ is the speed of light and $G$ is Newton's constant.
Thus, if the surface redshift and surface gravity can be constrained
separately, we can constrain $M$ and $R$.  

To do this requires fits to the data that (1)~are dramatically better
than Bose-Einstein fits, so that we have some confidence in the
inferences we draw from model atmosphere spectra, and (2)~distinguish
between compositions, surface gravities, and surface redshifts. Our
work on this program, which we are undertaking in collaboration with
Valery Suleimanov and Juri Poutanen, has yielded good initial results.
We find that when we fit the 102 contiguous 16-second segments of
data from the 4U~1820--30 superburst mentioned previously, assuming 
that the composition, surface gravity, and surface redshift remain the
same for all segments but that the surface radiative flux can change,
one example fit  of the Suleimanov et al. models gives 
$\chi^2/{\rm dof}=5394/5200$.
In contrast, the best Bose-Einstein joint fit to the
data, where we allow the temperature and chemical potential to vary
independently between segments, gives $\chi^2/{\rm dof}=5660/5100$.  This
comparison strongly favors the model atmosphere spectra, and there are
preliminary indications that composition, surface gravity, and 
surface redshift can be constrained.  However, we caution
that because the current models are known to make approximations
compared to the exact scattering kernel, any conclusions are premature
at this point.  Nonetheless, this approach seems promising.

In summary, we have recently performed the first comparison of
predicted spectra with the highest-precision data available from the
RXTE PCA. We found that although a Bose-Einstein spectrum fits all
individual segments well, previously published model atmosphere
spectra have shapes strongly inconsistent with the observed spectra.
This suggests caution in inferences made using these
spectra.  New spectral models provide promising descriptions of the
highest-precision data and may restrict the mass and radius via
constraints on the surface gravity and redshift, once they have been made
more accurate.

These results are based on research supported by NSF grant
AST0708424 at Maryland and by NSF grant AST0709015 and the
Fortner Chair at Illinois.

\end{document}